\documentclass[journal,12pt,onecolumn,draftclsnofoot,]{IEEEtran}

\usepackage[cmex10]{amsmath}
\usepackage[utf8]{inputenc}
\usepackage{multirow}
\usepackage{soul}
\usepackage{algorithm}
\usepackage{algpseudocode}
\usepackage{graphicx}
\usepackage{dsfont}
\usepackage{xcolor}
\usepackage{cite}

\usepackage{tabularray}
\usepackage{tabu}
\usepackage{ulem}

\usepackage[utf8]{inputenc}
\usepackage{mathtools}
\usepackage[mathscr]{euscript}
\usepackage{amsmath}
\usepackage{amssymb}
\usepackage{amsfonts}
\usepackage{amssymb}
\usepackage{amsmath}
\usepackage{verbatim}
\usepackage[T1]{fontenc}
\usepackage[utf8]{inputenc}
\usepackage{latexsym}
\usepackage{enumerate}
\usepackage{indentfirst}
\usepackage{calligra}
\usepackage{ulem}
\usepackage{graphicx}
\usepackage{array}
\usepackage{float}
\usepackage{bbm}

\usepackage{geometry}

\usepackage{mleftright}
\usepackage{colortbl}
\makeatletter
\let\old@ps@headings\ps@headings
\let\old@ps@IEEEtitlepagestyle\ps@IEEEtitlepagestyle
\def\psccfooter#1{%
 \def\ps@headings{%
 \old@ps@headings%
 \def\@oddfoot{\strut\hfill#1\hfill\strut}%
 \def\@evenfoot{\strut\hfill#1\hfill\strut}%
 }%
 \def\ps@IEEEtitlepagestyle{%
 \old@ps@IEEEtitlepagestyle%
 \def\@oddfoot{\strut\hfill#1\hfill\strut}%
 \def\@evenfoot{\strut\hfill#1\hfill\strut}%
 }%
 \ps@headings%
}
\makeatother

\usepackage{soul}
\usepackage{multirow}
\usepackage{soul,color}
\usepackage{graphicx}
\usepackage{dsfont}
\usepackage{subcaption}

\usepackage[utf8]{inputenc}
\usepackage{mathtools}
\usepackage[mathscr]{euscript}
\usepackage{amsmath}
\usepackage{amssymb}
\usepackage{amsfonts}
\usepackage{verbatim}
\usepackage[T1]{fontenc}
\usepackage[utf8]{inputenc}

\usepackage{latexsym}
\usepackage{enumerate}
\usepackage{indentfirst}
\usepackage{calligra}
\usepackage{ulem}
\usepackage{multirow}

\usepackage{array}
\usepackage{float}
\usepackage{bbm}

\usepackage{eurosym}

\usepackage{hyperref}

\usepackage{caption}
\usepackage{tikz}
\usepackage{pgfplots}

\pgfplotsset{compat=1.8}
\usepgfplotslibrary{statistics}
\pgfmathdeclarefunction{fpumod}{2}{%
 \pgfmathfloatdivide{#1}{#2}%
 \pgfmathfloatint{\pgfmathresult}%
 \pgfmathfloatmultiply{\pgfmathresult}{#2}%
 \pgfmathfloatsubtract{#1}{\pgfmathresult}%
 \pgfmathfloatifapproxequalrel{\pgfmathresult}{#2}{\def\pgfmathresult{5}}{}%
 }

\usepackage{tabularx}
\usepackage{flushend}
\usepackage{textcomp}
\usepackage{xcolor}
\usepackage{import}

\usepackage{csvsimple}
\usepackage{tabularx}
\usepackage{adjustbox}

\usetikzlibrary{trees,decorations,shadows}
\tikzset{level 1/.style={sibling angle=45,level distance=4mm}}
\usetikzlibrary{arrows.meta}
\usepackage{forest}
\usetikzlibrary{external}

\let\oldtikzexternalgetnextfilename\tikzexternalgetnextfilename \renewcommand{\tikzexternalgetnextfilename}[1]{\oldtikzexternalgetnextfilename{#1}\expandafter\tikzsetnextfilename\expandafter{#1}}

\usepackage{pgfplotstable}
\usepackage[outline]{contour}
\contourlength{1.2pt}

\pgfplotsset{compat=1.13} 

\usetikzlibrary{spy}
\usetikzlibrary{calc}
\usetikzlibrary{fadings}
\usetikzlibrary{patterns}
\usetikzlibrary{shadows}
\usetikzlibrary{mindmap}
\usetikzlibrary{backgrounds}
\usetikzlibrary{shapes.symbols}
\usetikzlibrary{shapes.multipart}
\usetikzlibrary{shapes.geometric}
\usetikzlibrary{automata,positioning}
\usetikzlibrary{decorations.fractals} 
\usetikzlibrary{decorations.markings}
\usetikzlibrary{decorations.pathreplacing}
\usetikzlibrary{decorations.pathmorphing}
\usepackage{svg}
 
\usepackage{bm}

\usepackage{nomencl}
\makenomenclature
\tikzset{edge from parent/.style={segment angle=10,draw}}

\tikzset{
 my rounded corners/.append style={rounded corners=2pt},
}

\usepackage{lipsum}
\usepackage{adjustbox}
\usetikzlibrary{positioning, arrows.meta}
\usetikzlibrary{shapes,arrows}
\usepackage[RPvoltages]{circuitikz}
\newcommand{\bushere}[2]{%
    coordinate(tmp)
    ++(0,1) node[above]{#1} edge[ultra thick] ++(0,-2)
    ++(0,-2) node[below]{#2}
    (tmp)
}

\usepackage{makecell}

\usepackage{pgfpages}

\def\BibTeX{{\rm B\kern-.05em{\sc i\kern-.025em b}\kern-.08em
 T\kern-.1667em\lower.7ex\hbox{E}\kern-.125emX}}

\renewcommand{\nomgroup}[1]{%
 \ifthenelse{\equal{#1}{O}}{\item[\textit{Operators}]}{%
 \ifthenelse{\equal{#1}{I}}{\item[\textit{Indices}]}{%
 \ifthenelse{\equal{#1}{A}}{\item[\textit{Acronyms}]}{%
 `\ifthenelse{\equal{#1}{V}}{\item[\textit{Variables and parameters}]}{}}}}}
\usepackage{scalerel}
\usetikzlibrary{svg.path}
\definecolor{orcidlogocol}{HTML}{A6CE39}
\tikzset{
 orcidlogo/.pic={
 \fill[orcidlogocol] svg{M256,128c0,70.7-57.3,128-128,128C57.3,256,0,198.7,0,128C0,57.3,57.3,0,128,0C198.7,0,256,57.3,256,128z};
 \fill[white] svg{M86.3,186.2H70.9V79.1h15.4v48.4V186.2z}
 svg{M108.9,79.1h41.6c39.6,0,57,28.3,57,53.6c0,27.5-21.5,53.6-56.8,53.6h-41.8V79.1z M124.3,172.4h24.5c34.9,0,42.9-26.5,42.9-39.7c0-21.5-13.7-39.7-43.7-39.7h-23.7V172.4z}
 svg{M88.7,56.8c0,5.5-4.5,10.1-10.1,10.1c-5.6,0-10.1-4.6-10.1-10.1c0-5.6,4.5-10.1,10.1-10.1C84.2,46.7,88.7,51.3,88.7,56.8z};
 }
}

\newcommand\orcidicon[1]{\href{https://orcid.org/#1}{\mbox{\scalerel*{ \begin{tikzpicture}[yscale=-1,transform shape]
 \pic{orcidlogo};
 \end{tikzpicture}
 }{|}}}}

\begin{document}
\IEEEoverridecommandlockouts
\IEEEpubid{\makebox[\columnwidth]{979-8-3503-9042-1/24/\$31.00 \copyright 2024 IEEE \hfill } 
\hspace{\columnsep}\makebox[\columnwidth]{\hfill }}
%
\title{\huge{Improved Physics-Informed Neural Network based AC Power Flow for Distribution Networks}}

\author{\IEEEauthorblockN{Victor Eeckhout*\orcidicon{0009-0008-5648-6309}, 
Hossein~Fani*\orcidicon{0000-0003-3781-7404}
Md~Umar~Hashmi*\orcidicon{0000-0002-0193-6703},
and~Geert~Deconinck*\orcidicon{0000-0002-2225-3987}}
 
 \IEEEauthorblockA{*\textit{KU Leuven \& EnergyVille},
Genk, Belgium}

 \IEEEauthorblockA{Email: eeckhout.victor@student.kuleuven.be, (hossein.fani, mdumar.hashmi, geert.deconinck)@kuleuven.be}
 }


\maketitle

\begin{abstract}
Power flow analysis plays a critical role in the control and operation of power systems. The high computational burden of traditional solution methods led to a shift towards data-driven approaches, exploiting the availability of digital metering data. However, data-driven approaches, such as deep learning, have not yet won the trust of operators as they are agnostic to the underlying physical model and have poor performances in regimes with limited observability. To address these challenges, this paper proposes a new, physics-informed model. More specifically, a novel physics-informed loss function is developed that can be used to train (deep) neural networks aimed at power flow simulation. The loss function is not only based on the theoretical AC power flow equations that govern the problem but also incorporates real physical line losses, resulting in higher loss accuracy and increased learning potential. The proposed model is used to train a Graph Neural Network (GNN) and is evaluated on a small 3-bus test case both against another physics-informed GNN that does not incorporate physical losses and against a model-free technique. The validation results show that the proposed model outperforms the conventional physics-informed network on all used performance metrics. Even more interesting is that the model shows strong prediction capabilities when tested on scenarios outside the training sample set, something that is a substantial deficiency of model-free techniques.

\end{abstract}

\begin{IEEEkeywords}
AC power flow, physics-informed neural network, model-free techniques, data-driven model, distribution networks.
\end{IEEEkeywords}



 \pagebreak

\tableofcontents

 \pagebreak

\section{Introduction}

AC power flow (ACPF) analysis aims at obtaining complete voltage angles and magnitudes for each bus in a power system, given specific loads, generator real power, and voltage conditions. \cite{saadat1999power}. It is an invaluable instrument for the system operator to maintain awareness of the current state of the system. This awareness is fundamental to their core responsibilities, which include planning, operation, economic scheduling, and the exchange of power between utilities \cite{pagnier2021physics}, \cite{nitve2014steady}. Given its non-linear nature, the power flow problem is traditionally solved using model-based numerical iterative techniques, including Newton-Raphson and Gauss-Seidel. However, as distribution systems differ from transmission networks in various aspects (including weakly meshed or radial topologies, feeder loading and high R/X ratios, and the change of system topology owing to reconfiguration application and self-healing capability), these methods result in ill-conditioned problems and slow convergence \cite{abbasi2023probabilistic}. Additionally, with the uninterrupted growth of connections and the increasing complexity of distribution systems, the computational burden of using traditional methods has grown too large. For this reason, alternative methods to solve the ACPF problem in the context of the distribution system have arisen.

A general classification of power flow approaches is based on whether an explicit physical model, in the form of the differential-algebraic ACPF equations and/or system topology information and network parameters, was used to obtain a solution. It considers, besides model-based methods, model-free (or data-driven) and hybrid methods \cite{hu2020physics}. Data-driven methods, such as machine learning (ML), fully omit the use of a physical model. Instead, they leverage available historical metering data to effectively establish the functional relationships between measurements and system states \cite{ngo2024physics}. For this reason, data-driven methods are also often referred to as `black-box' methods. 
Due to their strong adaptive learning and generalization capabilities, they are particularly useful when the relationships between input and target features are not well understood or when statistical patterns (e.g. introduced by increased DER penetration) are missed by physical relations alone \cite{zhu2019electric}. In the context of ACPF problems, ML methods are advantageous either when topological or structural network data is unavailable or the computational burden of the model is high. 
Despite some success in learning the physical relations of the system, black-box models can introduce excessive parameters and risk overfitting \cite{wu2023spatio}.
Exploiting prior physical knowledge of systems, a new hybrid paradigm called physics-informed methods, particularly physics-informed neural networks (PINNs), has gained significant traction in recent years for solving complex, non-linear systems. This approach incorporates physical laws as constraints during training, and its applicability has been extended to various domains, including power system problems like ACPF.

\vspace{-1pt}

Except for some studies, like \cite{wu2023spatio} and \cite{zamzam2020physics} in which the authors use the admittance matrix as graph filters in a graph neural network (GNN), most of the works concerning physics-informed power flow rely on the same structure to incorporate physics in the model. In these studies, the physical equations are incorporated as regularization constraints imposed on the model by incorporating them within the training loss function. Penalizing deviations from these constraints allows the networks to effectively choose the weights and bias parameters based on the underlying relationships. 
In \cite{nellikkath2022physics}, the authors insert a series of constraints, including the physical ACPF constraint, in order to solve the optimal ACPF. The different constraints are weighed, effectively introducing the weights as hyperparameters of the model. The authors of \cite{lopez2024optimal} and \cite{hu2020physics} use the same methodology, in which the former work uses both branch- and node-wise ACPF constraints to weigh, and the latter work proposes to separate the loss calculation by different multilayer perceptrons in an encoder-decoder fashion before adding and weighing them. The authors in \cite{ngo2024physics} use only the ACPF constraint besides mean square error (or MSE) loss, weighing both losses equally. Another type of learning introduces the physical constraint in a regularization term to prevent overfitting and model the system's complexity \cite{wang2020comprehensive}. 
This is done in \cite{ostrometzky2019physics}, in which the ACPF constraint serves as the only regularization term. The study in \cite{jalving2024physics} uses multiple regularization terms under the same coefficient, in which this coefficient is optimized during each epoch using dual-Lagrange optimization. In \cite{lei2020data}, the authors propose a similar structure but use stacked extreme learning machines (SELMs), in which branch flows and complex voltage learning are done in different hidden layers. On the other hand, some works only consider the physical ACPF constraint in their training loss, resulting in a pure physics-informed model. This is done, amongst others, in \cite{pagnier2021physics} and in \cite{donon2020neural}, in which the latter work introduces a discount factor as a new hyperparameter to emphasize later corrections of the loss function. Lastly, \cite{lin2023powerflownet} considers both, separately, pure physics-informed, MSE and mixed loss functions.

This work aims to benchmark only the physical component of the power flow prediction objective,
to measure and benchmark the value of inserting physics into the model. For this, a novel, improved physical ACPF loss function is developed that incorporates physical line losses, accounting for the real behavior of power flow in distribution networks. This loss function will be incorporated into a GNN, which is then eventually benchmarked against both a conventional physics-informed GNN and a classic model-free network.

\subsection{Contributions of the paper}
The contributions of the paper are as follows:\\
$\bullet~$\textit{Proposing a new physics-informed loss function}:
    This work proposes a novel physics-informed loss function to train machine learning algorithms used for power flow simulation. The loss function is based on the node-wise power flow imbalance and is the first work, to the best of the authors' knowledge, that takes into account the real line losses caused by heat dissipation due to the Joule effect.\\
$\bullet~$\textit{Quantifying the limitations of model-free techniques}:
    The paper shows that classic model-free techniques have low to no prediction capabilities regarding test scenarios outside the sample space. On the other hand, models trained with the proposed loss function show promising results even for scenarios it has not seen. This observation clearly shows that the model was able to learn the underlying physics based on the new loss function.\\
$\bullet~$\textit{Numerical evaluation}: Benchmarking the proposed solution with model-free and model-based ACPF for a 3-bus system.









The rest of the paper is organized as follows:
Section \ref{section2} details the proposed loss function used for tuning ACPF model.
Section \ref{section3} describes the performance metrics used for evaluating the numerical results.
Section \ref{section4} details the numerical results.
Section \ref{section5} concludes the paper.

 \pagebreak

\section{Proposed loss function}
\label{section2}
\begin{figure}[htb]
    \centering
    \adjustbox{width=0.8\textwidth}{\tikzset{font={\fontsize{12pt}{14pt}\selectfont}}
\begin{tikzpicture}[semithick]
  \draw (0,0) node[vsourcesinshape, rotate=90](S){} (S.south) -- ++(1,0)
  \bushere{$\mathbf{V}_0 = 1\angle0$}{Bus 0} -- ++(1,0)
  to[oosourcetrans] ++(1,0)
  to[generic, l_={$y_{01}=G_{01}+jB_{01}$}] ++(3,0)
  \bushere{$\mathbf{V}_1 = V_{1}\angle\theta_1$}{Bus 1} -- ++(0.5,0)
  to[generic, l_={$y_{12}=G_{12}+jB_{12}$}] ++(3.5,0)
  \bushere{$\mathbf{V}_2 = V_{2}\angle\theta_2$}{Bus 2} -- ++(1,0)
  to[generic, l_={$y_{23}=G_{23}+jB_{23}$}] ++(3.5,0)
  \bushere{$\mathbf{V}_3 = V_{3}\angle\theta_3$}{Bus 3}
      ;
  \draw (7.5,0.5) node[anchor=west] {$P_1$};
  \draw[ultra thick, ->] (6.4,0.5) -- (7.5,0.5);
  \draw (11.5,0.5) node[anchor=west] {$P_2$};
  \draw[ultra thick, ->] (10.4,0.5) -- (11.5,0.5);
  \draw (16,0.5) node[anchor=west] {$P_3$};
  \draw[ultra thick, ->] (14.9,0.5) -- (16,0.5);

  \draw (4,-1.5) node[anchor=west] {$P_{10}$};
  \draw[ultra thick, ->] (3,-1.5) -- (4,-1.5);
  \draw (3,-2) node[anchor=east] {$P_{01}-\Delta P^{loss}_{01}$};
  \draw[ultra thick, ->] (4,-2) -- (3,-2);

  \draw (9,-1.5) node[anchor=west] {$P_{21}$};
  \draw[ultra thick, ->] (8,-1.5) -- (9,-1.5);
  \draw (8,-2) node[anchor=east] {$P_{12}-\Delta P^{loss}_{12}$};
  \draw[ultra thick, ->] (9,-2) -- (8,-2);

  \draw (13.5,-1.5) node[anchor=west] {$P_{32}$};
  \draw[ultra thick, ->] (12.5,-1.5) -- (13.5,-1.5);
  \draw (12.5,-2) node[anchor=east] {$P_{23}-\Delta P^{loss}_{23}$};
  \draw[ultra thick, ->] (13.5,-2) -- (12.5,-2);

\end{tikzpicture}}
    \caption{3-bus network}
    \label{fig:3_bus_feeder_with_losses}
    \vspace{-4pt}
\end{figure}
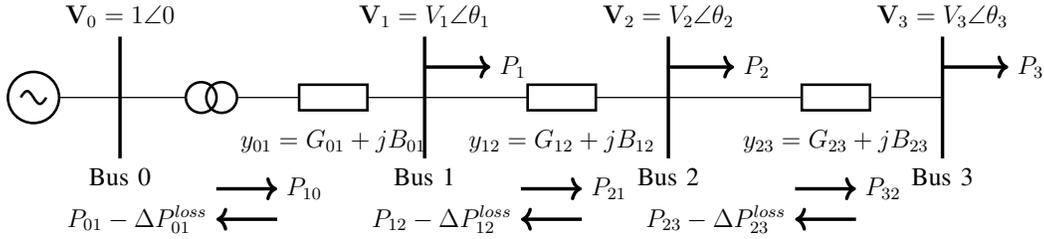

This section introduces the Alternating Current power flow (ACPF) equations that mathematically describe the physical power flow problem. As these equations holistically capture the physics of the problem, it is not surprising they will form the basis for the proposed physics-informed power flow solver framework, presented in this section. In this work, all derivations and implementations are intended for applications in single-phase radial distribution networks. Extrapolating the methodology to multi-phase distribution networks will be the subject of future work. Fig. \ref{fig:3_bus_feeder_with_losses} shows the notation used in this paper.


\subsection{AC Power Flow}
\label{sec:ACPF}
Consider a distribution network that operates over a grid graph $\mathcal{G}=(\mathcal{N},\mathcal{E})$, where $\mathcal{N}$ represents the set of nodes or buses (which can be generator or load buses) in the system with $\lvert \mathcal{N}\rvert=N$, and $\mathcal{E}=(i,j)$ represents the set of lines in the system, connecting the buses $i$ and $j$ \cite{pagnier2021physics}. The equations that mathematically govern the steady redistribution of AC power in this system, are called the AC Power Flow equations. This paper uses the polar form of the ACPF equations, which are based on the equations expressing the flow of power through a distribution line:
\begin{align}
    P_{ij} &= G_{ij}\left( V_iV_j\cos(\theta_i-\theta_j)-V_i^2\right)+B_{ij}V_iV_j\sin(\theta_i-\theta_j) \label{eq:active_theo_lineflow}\\
    Q_{ij} &= -B_{ij}\left( V_iV_j\cos(\theta_i-\theta_j)-V_i^2\right)+G_{ij}V_iV_j\sin(\theta_i-\theta_j)
\end{align}

In the equations above, $V_i$ and $\theta_i$ denote respectively the voltage magnitude and angle, measured at node $i$. Additionally, $G_{ij}$ and $B_{ij}$ represent the conductance and susceptance of the line between buses $i$ and $j$. These equations for the active and reactive line flows allow us to construct ACPF equations (\ref{eq:KCL_P}) and (\ref{eq:KCL_Q}), by considering the power balance at each node. This power balance arises from Kirchoff's current law, which states that the net consumption (or negative injection) at a certain node $i$ equals the line flows arriving at this node from all its neighbouring nodes. In Equations (\ref{eq:KCL_P}) and (\ref{eq:KCL_Q}), $\mathcal{N}(i)$ represents the set of neighbouring buses or nodes of bus $i$, and $\theta_{ij}=\theta_i-\theta_j$ is the difference between the voltage angles at bus $i$ and $j$.
\begin{align}
  P_i  &= \sum_{j\in \mathcal{N}(i)}P_{ij} = \sum_{j\in \mathcal{N}(i)} V_iV_j\left(G_{ij}\cos\theta_{ij}+B_{ij}\sin\theta_{ij}\right) \label{eq:KCL_P}\\
  Q_i &= \sum_{j\in \mathcal{N}(i)}Q_{ij} = \sum_{j\in \mathcal{N}(i)} V_iV_j\left(G_{ij}\sin\theta_{ij}-B_{ij}\cos\theta_{ij}\right) \label{eq:KCL_Q}
\end{align}

In the classic power flow problem, the known variables at each generator bus are the active power and the voltage magnitude, while the voltage angle and reactive power injection remain unknown. On the other hand, at each load bus, both the active and reactive power injection are known, while the voltage magnitude and angle remain unknown. Finally, at the reference bus the voltage magnitude and angle are known and generally assumed to be $\mathbf{V}_0 = 1\angle 0^{\circ}$. When only solving for the voltage angle at P-V buses, the amount of unknowns in the power flow system equals $2(N-1)-(R-1)$ where $R\leq N$ is the number of generator buses in the system. When there are only load buses in the system (and thus $R=1$, only the reference bus), the amount of unknowns simplifies to $2(N-1)$. This equals the number of available equations introduced by the system in Equations (\ref{eq:KCL_P}) and (\ref{eq:KCL_Q}). Due to the presence of the trigonometric entities, the system is non-linear, meaning that it needs to be solved with numerical and iterative methods.


\subsection{Proposed physics-informed loss function}
\label{subsec:proposed_loss_function}

As stated before, recently, machine learning methods have emerged, leveraging historical data to simulate Power Flow. They are used to learn the described relationships between the known power values (also called the input features) and unknown voltages (also called the output features, targets or labels) at each bus. Mathematically, the input features $\mathbf{x}^i$ can be represented as (row) vectors consisting of the power measured at each bus in $\mathcal{N}$ for each time-step $i$, effectively constituting a time series. 
In this work, we construct the physics-informed ACPF model using graph neural networks, assuming only active power load and injection in radial distribution networks, therefore no reactive power inputs are assumed (equivalent to a load angle $\delta=0^{\circ}$). 
Further, study is needed to include reactive power in the proposed ACPF model.
Adding these input feature vectors as the rows of a matrix results in the input-feature matrix $X$:
\begin{equation}
    X = 
    \begin{bmatrix}
        \mathbf{x}^1 \\
        \vdots\\
        \mathbf{x}^T
    \end{bmatrix}
    =
    \begin{bmatrix}
        \mathbf{P_1} & \cdots & \mathbf{P_N}    
    \end{bmatrix} 
    = 
    \begin{bmatrix}
        P^1_1 & \cdots & P^1_N\\
        \vdots & \ddots & \vdots\\
        P^T_1 & \cdots & P^T_N\\
    \end{bmatrix} \in \mathbb{R}^{T\times N}
\end{equation}

On the other hand, the output features $\mathbf{y}^i$ are again (row) vectors for each time-step $i$, but this time, they consist of the voltage magnitudes and angles for each bus in $\mathcal{N}$, shaping the output feature matrix $Y$:
\begin{align}
    Y &=
    \begin{bmatrix}
        \mathbf{y}^1 \\
        \vdots\\
        \mathbf{y}^T
    \end{bmatrix}
    =
    \left[
    \begin{array}{ccc|ccc}
        V^1_1 & \cdots & V^1_N & \theta^1_1 & \cdots & \theta^1_N \\
        \vdots & \ddots & \vdots  & \vdots & \ddots & \vdots \\
        V^T_1 & \cdots & V^T_N & \theta^T_1 & \cdots & \theta^T_N 
    \end{array}
    \right] \in \mathbb{R}^{T\times 2N}
\end{align}

As every input feature vector $\mathbf{x}^i$ in the training set is linked to an output feature vector $\mathbf{y}^i$ for all $i$, the ACPF problem can be treated as a supervised learning problem within machine learning. 
The goal of a supervised learning problem is to learn a model, i.e., a function or mapping $f(\mathbf{x})$ aimed at predicting the value of $\mathbf{y}$ for a new sample $\mathbf{x}$, based on a set of input-target feature pairs $\{(\mathbf{(x}^i,\mathbf{y}^i)\}_{i=1}^T$ \cite{nie2018investigation}. As in the ACPF problem, this model is aimed to return a continuous rather than a discrete target value for $\mathbf{x}$, it is considered a regression rather than a classification problem.  This regression problem can be mathematically formulated as:


\begin{equation}
    \label{eq:ML_algo}
    \underset{f(\mathbf{x})}{\min}\frac{1}{T}\sum^T_{i=1}L_{\theta}(f(\mathbf{x}^i)-\mathbf{y}^i)
\end{equation}

Equation (\ref{eq:ML_algo}) tries to find the model $f$ that, based on the (training) input features and labels, minimizes a certain loss (or cost) function $L_\theta$. The most common choice for this loss function is, $L_{MSE}$, the `Mean Squared Error' (or MSE). However, in the ACPF problem, the physical model is established and bounded by Equations (\ref{eq:KCL_P}) and (\ref{eq:KCL_Q}), eliminating the necessity to let a `black-box' method learn the underlying power flow equations. Therefore, instead of training (any) model to imitate the output of a power flow solver by minimizing the general MSE loss function $L_{MSE}$, this work proposes a novel physics-informed loss function. This novel loss function $L_{phys}$ is intended to harness the physical constraints of the ACPF problem by directly penalizing violations of Kirchoff's laws that govern the problem. This effectively creates a hybrid model at the interface between data-driven and model-based solutions.

As Equations (\ref{eq:KCL_P}) and (\ref{eq:KCL_Q}) incorporate the invariants and symmetries of the ACPF problem, they are taken as a starting point to derive the physical loss function. An evident way to derive a loss function from these equations, similar to as in \cite{lin2023powerflownet}, is by deriving the theoretical power imbalance at each node $i$. This is equal to the difference between the left- and right-hand sides in (\ref{eq:KCL_P}) and (\ref{eq:KCL_Q}). Considering only the active power portion results in the following loss function:
\begin{equation}
    \label{eq:phys_loss_theo}
        L_{phys} = \frac{1}{\left|N\right|}\sum_{i\in N}||{\Delta P_i}||^2_2, ~\text{with }\Delta P_i = P_i - \sum_{j\in \mathcal{N}(i)}P_{ij}
\end{equation}

The proposed loss function will deviate from the model in Equation (\ref{eq:phys_loss_theo}) in the following ways. Firstly, while $\Delta P_i$ in (\ref{eq:phys_loss_theo}) should in theory converge to zero, it is inaccurate in real circumstances. This is mainly because $P_{ij}$ as defined in Equation (\ref{eq:active_theo_lineflow}), is the theoretical flow of active power in the line connecting bus $i$ and $j$. In reality, the power flows suffer from a loss component $\Delta P^{loss}_{ij}$ ($=\Delta P^{loss}_{ji}$) introduced by heat dissipation in the lines due to the Joule effect. This effect is prominent in distribution networks, as the $R/X$ values are generally large. To arrive at the proposed $L_{phys}$, therefore, the following extension to Equation (\ref{eq:active_theo_lineflow}) is introduced:
\begin{equation}
    P_{ij} \approx -(P_{ji}-\Delta P^{loss}_{ij})\label{eq:active_lineflow_real}
\end{equation}
$\Delta P^{loss}_{ij}$ is approximated according to Joule's law with \cite{li2022numerical}:
\begin{equation}
    \label{eq:joule_losses}
  \Delta P^{loss}_{ij} \approx G_{ij}(V_i^2+V_j^2-2V_iV_j\cos(\theta_i-\theta_j))
\end{equation}
Combining (\ref{eq:active_theo_lineflow}), (\ref{eq:phys_loss_theo}), (\ref{eq:active_lineflow_real}) and (\ref{eq:joule_losses}) results in the proposed physical loss function.

\begin{figure}[htb]
  \centering
  \adjustbox{width=0.7\textwidth}{\begin{tikzpicture}[semithick]
  \draw (0,0) -- ++(3,0)
      \bushere{$\mathbf{V}_1 = V_{1}\angle\theta_1$}{Bus 1}
      ;
  \draw (4,0.5) node[anchor=west] {$\sum_{i\neq 0} P_i$};
  \draw[ultra thick, ->] (3,0.5) -- (4,0.5);

  \draw (1.0,-0.5) node[anchor=east] {$P_0-\sum \Delta P_{ij}$};
  \draw[ultra thick, ->] (2.0,-0.5) -- (1.0,-0.5);

\end{tikzpicture}}
  \caption{Approximated power imbalance at bus 1}
  \label{fig:approximated_power_imbalance_bus_1}
  \vspace{-4pt}
\end{figure}
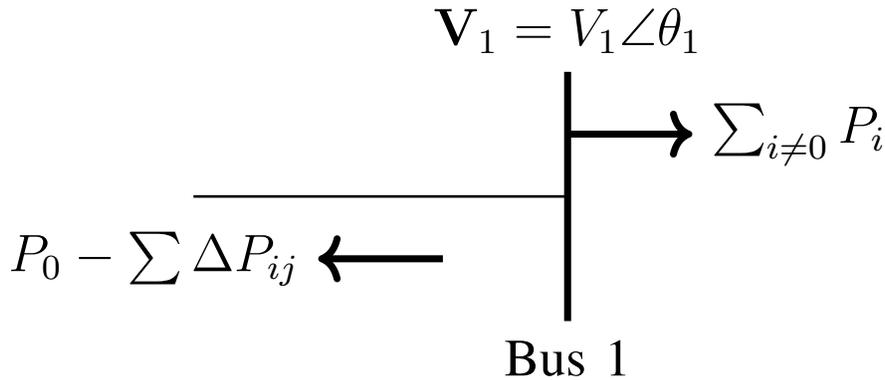

Due to practical considerations, a final adaptation is necessary. More specifically, computing the proposed physical imbalance at the bus connected directly to the slack bus ($\Delta P_1$) according to (\ref{eq:phys_loss_theo}) is difficult, mainly because Equation (\ref{eq:active_theo_lineflow}) is inaccurate for the flow from the slack bus to the first bus $P_{10}$. 
Therefore, an approximation for $\Delta P_1$ is used, obtained from the fact that the total power drawn at the transformer ($P_0$; here defined as negative) should equal the sum of the loads and the sum of the losses in the lines (see Figure \ref{fig:approximated_power_imbalance_bus_1}):


\begin{equation}
  \Delta P_1 \approx \sum_{i} P_i - \sum \Delta P^{loss}_{ij}
\end{equation}

\vspace{-10pt}

 \pagebreak

\section{Performance metrics used}
\label{section3}
The use of machine learning methods like deep neural networks (DNNs) to solve the ACPF problem is attractive mainly because of the scalability advantages. Predicting the voltages with a trained DNN is of much lower time complexity than actually solving the non-linear ACPF system with traditional model-based methods. However, this gain in computation time comes at the cost of reduced accuracy compared to these methods. Therefore, the evaluation of the models will focus on accuracy. More specifically, the focus will be skewed towards the accuracy of the model's predictions on out-of-sample test data. Therefore, next to classical key performance indicators (KPIs), like RMSE, new indicators to evaluate this out-of-sample behaviour, like `Overvoltage Accuracy' and `False Positive Rate', will be defined below for a set of $T$ prediction-target voltage test values $\{(\hat{y},y)\}^T_{i=1}$:


\begin{enumerate}
    \item \textit{Root Mean Square Error (RMSE)}: 
    The RMSE as defined in Equation (\ref{eq:RMSE}) is one of the two main accuracy indicators of a regression model. It is a measure of the accuracy of the predictions, aggregated over the whole test set.
    \begin{equation}
        \label{eq:RMSE}
        RMSE = \sqrt{\frac{1}{T}\sum_{i=0}^T(y_i-\hat{y}_i)^2}
    \end{equation}
    
    \item \textit{Pearson Correlation Coefficient (PCC)}:
    The Pearson correlation coefficient is a correlation metric that can be seen as a proxy for how well the prediction profile emulates the actual voltage profile. The correlation is again measured over the whole test set and defined as follows: 
    \begin{equation}
        \label{eq:PCC}
        \frac{\sum ^n _{i=1}(\hat{y}_i - \bar{\hat{y}})(y_i - \bar{y})}{\sqrt{\sum ^n _{i=1}(\hat{y}_i - \bar{\hat{y}})^2} \sqrt{\sum ^n _{i=1}(y_i - \bar{y})^2}}
    \end{equation}

    \item \textit{Overvoltage Accuracy (OVA)}: 
    A metric used to evaluate the out-of-sample prediction capacity of the models is the overvoltage accuracy, which is defined as in Equation (\ref{eq:overvoltage_accuracy}). It is the ratio between the number of correctly predicted overvoltages, and the total number of overvoltages occurring in the test set. In other words, the OVA measures the type I error ($\alpha$), commonly used in statistics. A model with high overvoltage accuracy can predict peak behaviour particularly well.
      \begin{equation}
        \label{eq:overvoltage_accuracy}
        OVA = \frac{\lvert \{(y_i,\hat{y}_i)|y_i\geq V_{max}, \hat{y}_i\geq V_{max} \}\rvert}{\lvert \{(y_i,\hat{y}_i)|y_i\geq V_{max} \}\rvert}
      \end{equation}

    \item \textit{False Positive Rate (FPR)}:
    Correspondingly, another metric is introduced, this time to measure the type II error ($\beta$) of the model's predictions. The False Positive Rate is the ratio of the instances in which the model predicts an overvoltage while it does not occur in reality. The formal definition is given by Equation (\ref{eq:FPR}).
      \begin{equation}
        \label{eq:FPR}
        FPR = \frac{\lvert \{(y_i,\hat{y}_i)|y_i< V_{max}, \hat{y}_i\geq V_{max} \}\rvert}{\lvert \{(y_i,\hat{y}_i)\}\rvert}
      \end{equation}
\end{enumerate}

 \pagebreak

\section{Numerical evaluation}
\label{section4}
In this section, the proposed model of Section \ref{subsec:proposed_loss_function} for the novel physics-informed loss function to train a deep neural network is numerically evaluated. 
 This proposed model, referred to as `GNNp', is benchmarked with another physics-informed neural network-based ACPF model proposed by Lin et al. in \cite{lin2023powerflownet} and a model-free ACPF model based on Extreme Gradient Tree Boosting Regressor introduced in \cite{chen2016xgboost}.
 The `PowerFlowNet' model is trained with their physics-informed loss function in \cite{lin2023powerflownet} that does not incorporate physical losses. This model is referred to as `GNNb'.
 The model-free ACPF model is referred to as `XGB'.
The numerical case study performed uses the three models to simulate the AC Power Flow in a simple 3-bus test network. First, this network is introduced, and next the results of their voltage profile predictions are evaluated on the KPIs detailed in Section \ref{section3}.

\subsection{Network under review: 3-bus radial feeder}

To test the proposed model, a minimal single-phase radial feeder is used. The feeder, which represents a small distribution network, consists of three load buses that are all connected to the main and only branch. This branch is connected to the slack bus, bus 0, through a $20$ kVA rated $1-$phase transformer. This 3-bus network is depicted in Figure \ref{fig:3_bus_feeder_with_losses} and line parameters are taken from \cite{UmarPaper}. Notice also the active line flows and losses according to (\ref{eq:active_lineflow_real}) that are indicated in the figure. The choice of the minimal topology is due to the consideration that both the physics needed to build the proposed loss function and the interpretability of the results are more easily verified at this small scale. Nonetheless, all derivations and considerations made are valid for larger-scale distribution systems.



\subsection{Scenarios for training and testing ACPF models}

\textit{Need for evaluating out-of-sample dataset for trained ACPF models}:
The load buses can be thought of as houses in a small neighbourhood for which the active power consumption per household, as measured by metering infrastructure, is available in a time series with hourly intervals. While it cannot be derived from these active power measurements directly, the possibility exists that a household features a photovoltaic (PV) installation that allows them to generate electricity and inject it into the network in case of a surplus. However, excessive injections can lead to rising voltages in the feeder, possibly damaging the network. Therefore, it would be interesting to see whether the proposed physics-informed neural network can learn the physics of these injections, to predict future overvoltages and avoid damage. These kinds of prediction capabilities are very valuable in applications like PV Hosting Capacity. Hosting capacity refers to the maximum amount of PV that can be connected to the distribution network without hindering the reliable operation of the distribution network \cite{koirala2022decoupled}.

To evaluate the (overvoltage) prediction abilities of the proposed physics-informed model (and its benchmarks), it is subjected to the following experiment. Each model (XGB, GNNb and GNNp) will be trained three times, each time with a different training set (TS1-TS3, see Tab. \ref{tab:training_cases_statistics}). In these training sets, the PV injection level is set to be kept within a similar range. This is done by using only two parameters. First, the available active power from the solar panels at the inverter is uniformly but randomly chosen within the interval $[1,7.5]kW_p$. Thus, abstraction is made of the difference in PV installations and incoming irradiances, similar to \cite{hashmi2023robust, hashmi2024analyzing}. The second parameter is the PV penetration in the distribution grid, this is equivalent to the number of households that have a PV installation available. In the different training sets, the number of installations is kept higher than two out of three consumers, resulting in a $[66,100]\%$ penetration rate range. Besides the PV injection properties of the training sets, they are also characterized by different set lengths $T$. This length represents the memory capacity of the metering infrastructure at the households. The smallest training set, TS1 has a length of 7 days of 24 data points per day (corresponding to a data memory of 1 week). Similarly, TS2 and TS3 have a length of respectively 30 days (or one month) and 365 days (or 1 year), see Tab. \ref{tab:training_cases_statistics}. Using OpenDSS, a deterministic power flow solver software, the voltage time series (or training target features) can be computed from the active power series (or input training features) to complete the training sets. The statistical characteristics associated with training sets are shown in Table \ref{tab:training_cases_statistics}. 
Note that the maximum observed voltage is around $1.081$ per unit on a $230~V$ voltage base for each of the training sets. 


\vspace{-4pt}
\begin{table}[!htp]
    \centering
    \setcellgapes{2pt}
    \makegapedcells
    \caption{Statistical characteristics of the training sets}
    \vspace{-5pt}
    \label{tab:training_cases_statistics}
    \resizebox{0.95\columnwidth}{!}{%
    \begin{tabular}{|p{2em}|p{9em}|p{2em}|p{2em}|p{2em}|p{2em}|p{4.4em}|}
        \hline
         Case & Length training set $T$ & $\mu$ & $\sigma$ & min & max& $V_i>1.05$ \\\hline
         TS1 & $T_1 = 7\times 24$ & 0.989 & 0.024 & 0.900 & 1.073 & 8/504\\
         TS2 & $T_2 = 30\times 24$ & 0.991 & 0.025 & 0.901 & 1.072 & 27/2160\\
         TS3 & $T_3 = 365\times 24$ & 0.986 & 0.023 & 0.862 & 1.081 & 124/26280\\
         \hline
    \end{tabular}
    }
\end{table}


\subsubsection{Out of sample scenarios for ACPF testing}
Next, the models, trained with the different training sets are tested on two different test cases C1 and C2, both with a length of 30 days. The two test cases are created such that the characteristics (especially those regarding the maximum and amount of overvoltages) of the first one resemble those of the training sets, and those of the second one are more extreme. The voltage level in the second test case is raised by increasing the interval of generated active power to $[5,12]~kW_p$, which with the same PV penetration interval raises the average total PV injection. The characteristics of both test sets C1 and C2 are shown in Table \ref{tab:test_cases_statistics}.

\begin{table}[!h]
    \centering
    \setcellgapes{2pt}
    \makegapedcells
    \caption{Statistical characteristics of the different test cases}
    \label{tab:test_cases_statistics}
    \resizebox{0.95\columnwidth}{!}{%
    \begin{tabular}{|p{1.5em}|p{3.7em}|p{2.6em}|p{2em}|p{2em}|p{2em}|p{2em}|p{4.4em}|p{4.4em}|}
        \hline
         Case & penetr.[\%] & $kW_p$ & $\mu$ & $\sigma$ & min & max & $ V_i>1.05$ & $V_i>1.08$\\\hline
         C1 & $[66,100]$ & $[1,7.5]$ & 0.986 & 0.022 & 0.878 & 1.071 & 6/2160 & 0/2160\\
         C2 & $[66,100]$ & $[5,12]$ & 0.999 & 0.035 & 0.898 & 1.179 & 165/2160 & 59/2160\\
         \hline
    \end{tabular}}
\end{table}
\vspace{-10pt}

\subsection{Numerical results}

The prediction results of all the models are grouped per training set on which they were trained and are displayed in Tables \ref{tab:results_training_set_1}, \ref{tab:results_training_set_2} and \ref{tab:results_training_set_3} for TS1, TS2 and TS3, respectively. For the calculations of the Overvoltage Accuracy and False Positive Rate, an overvoltage threshold of $V_{max}=1.08~p.u.$ is chosen, this is not coincidentally around the maximum observed overvoltage values within TS1-TS3 (see again Tab. \ref{tab:training_cases_statistics}).

\begin{table}[!h]
    \centering
    \setcellgapes{1pt}
    \makegapedcells
    \caption{Training set 1 (TS1)}
    \vspace{-5pt}
    \label{tab:results_training_set_1}
    \resizebox{0.95\columnwidth}{!}{
    \begin{tabular}{|p{3.5em}|p{2.5em}|p{2.5em}|p{2.7em}|p{5.5em}|p{5.5em}|p{2.5em}|}
        \hline
        \multirow{2}{4em}{Test case} & \multirow{2}{*}{Model} & \multicolumn{5}{c|}{Metrics}\\
        \cline{3-7}
        & &RMSE [V] & PCC [\%] & OVA [\%] & OVA at bus 3 [\%] & FPR [\%]\\
        \hline
        \multirow{3}{*}{C1}& XGB & 0.94 & 98.31 & nan (0/0) & nan (0/0) & 0.00\\
        & GNNb & 14.99 & -4.82 & nan (0/0) & nan (0/0) & 2.92\\
        & GNNp & 14.24 & 20.40 & nan (0/0) & nan (0/0) & 7.27\\
        \hline
        \multirow{3}{*}{C2}& XGB & 2.46 & 96.47 & 0.00 (0/57)& 0.00 (0/57)& 0.00\\
        & GNNb & 19.02 & -18.62 & 6.78 (4/59)& 5.36 (3/57)& 3.56\\
        & GNNp & 17.21 & 21.22 & 33.90 (20/59)& 33.33 (19/57) & 10.00\\
        \hline
    \end{tabular}
    }
\end{table}

\begin{table}[!h]
    \centering
    \setcellgapes{1pt}
    \makegapedcells
    \caption{Training set 2 (TS2)}
    \label{tab:results_training_set_2}
    \vspace{-5pt}
    \resizebox{0.95\columnwidth}{!}{
    \begin{tabular}{|p{3.5em}|p{2.5em}|p{2.5em}|p{2.7em}|p{5.5em}|p{5.5em}|p{2.5em}|}
        \hline
        \multirow{2}{4em}{Test case} & \multirow{2}{*}{Model} & \multicolumn{5}{c|}{Metrics}\\
        \cline{3-7}
        & &RMSE [V] & PCC [\%] & OVA [\%] &  OVA at bus 3 [\%] & FPR [\%]\\
        \hline
        \multirow{3}{*}{C1}& XGB & 0.68 & 99.13 & nan (0/0)& nan (0/0)& 0.00\\
        & GNNb & 10.88 & -27.74 & nan (0/0)& nan (0/0)& 1.25\\
        & GNNp & 3.02 & 80.99 & nan (0/0)& nan (0/0)& 0.09\\
        \hline
        \multirow{3}{*}{C2}& XGB & 2.38 & 96.75 & 0.00 (0/59)& 0.00 (0/57)& 0.00\\
        & GNNb & 15.07 & -36.92 & 0.00 (0/59)& 0.00 (0/57)& 1.16\\
        & GNNp & 4.96 & 80.45 & 50.85 (30/59)& 49.12 (28/57) & 0.56\\
        \hline
    \end{tabular}
    }
\end{table}

\begin{table}[!h]
    \centering
    \setcellgapes{1pt}
    \makegapedcells
    \caption{Training set 3 (TS3)}
    \vspace{-5pt}
    \label{tab:results_training_set_3}
    \resizebox{0.94\columnwidth}{!}{
    \begin{tabular}{|p{3.5em}|p{2.5em}|p{2.5em}|p{2.7em}|p{5.5em}|p{5.5em}|p{2.5em}|}
        \hline
        \multirow{2}{4em}{Test case} & \multirow{2}{*}{Model} & \multicolumn{5}{c|}{Metrics}\\
        \cline{3-7}
        & &RMSE [V] & PCC [\%] & OVA [\%] & OVA at bus 3 [\%] & FPR [\%]\\
        \hline
        \multirow{3}{*}{C1}& XGB & 0.19 & 99.93 & nan (0/0) & nan (0/0) & 0.00\\
        & GNNb & 12.90 & -32.83 & nan (0/0)& nan (0/0) & 5.05\\
        & GNNp & 2.25 & 90.94 & nan (0/0)& nan (0/0) & 0.05\\
        \hline
        \multirow{3}{*}{C2}& XGB & 1.74 & 98.24 & 8.47 (5/59)& 8.77 (5/57)& 0.00\\
        & GNNb & 17.49 & -56.12 & 3.39 (2/59)& 3.51 (2/57)& 3.43\\
        & GNNp & 3.10 & 92.15 & 52.54 (31/59)& 52.63 (30/57)& 0.56\\
        \hline
    \end{tabular}}
    \vspace{-10pt}
\end{table}

\begin{figure}[bp]
  \centering
  \includegraphics[width=0.76\linewidth]{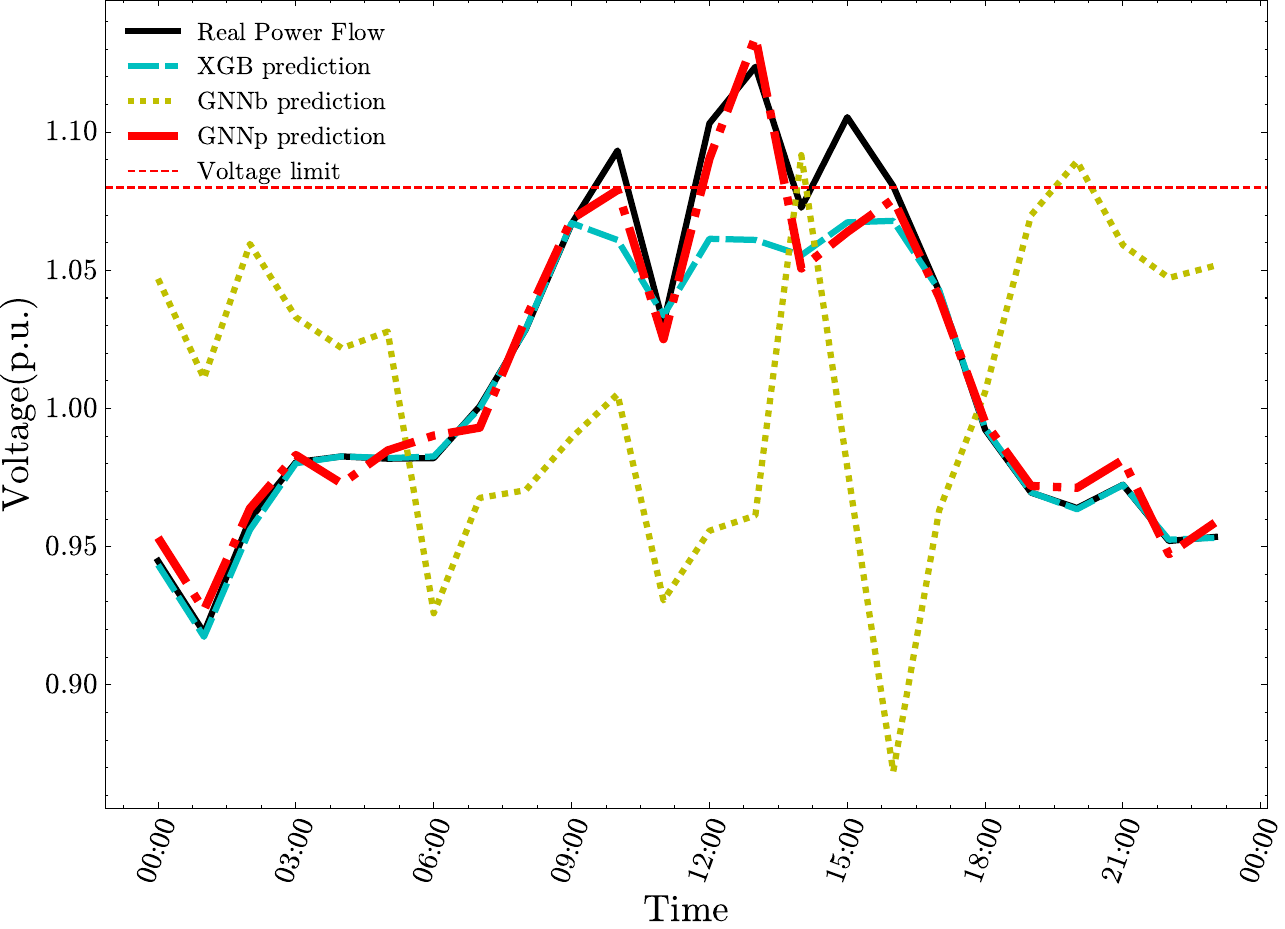}
  \vspace{-4pt}
  \caption{Voltage profiles for 1 day of the test data comparing models: model-free XGB, and physics-informed GNNb and GNNp.}
  \label{fig:voltage}
\end{figure}

The length of the training set clearly affects the model evaluation performance metrics. From Tables \ref{tab:results_training_set_1}, \ref{tab:results_training_set_2} and \ref{tab:results_training_set_3}, note that RMSE and FPR decreases for TS1 to TS3, while PCC and OVA increases for TS1 to TS3.
Although, the PCC for the XGB model outperforms others, note how XGB fails to predict over-voltage instances. This is also visualized in Fig. \ref{fig:voltage}. The primary reason why the model-free XGB ACPF model is not able to predict over-voltage instances accurately is because in the training dataset, such violations were not observed. This is a key limitation of model-free ACPF techniques. On the other hand, the proposed model, GNNp, outperforms all benchmarking models in predicting OVA. However, while the prediction for TS2 and TS3 still lies around 50\%, indicating the underlying physics was fairly accurately learned by the trained model, the accuracy for OVA for the proposed GNNp can be improved with better scenario generation, which in effect would lead to accurate learning of underlying physics.
For the FPR, GNNp is substantially better than the benchmarked GNNb. The proposed model also outperforms GNNb for PCC, RMSE and OVA.

\begin{figure}[htb]
    \centering
    \begin{subfigure}[b]{0.99\textwidth}
        \centering
        \includegraphics[width=0.8\textwidth]{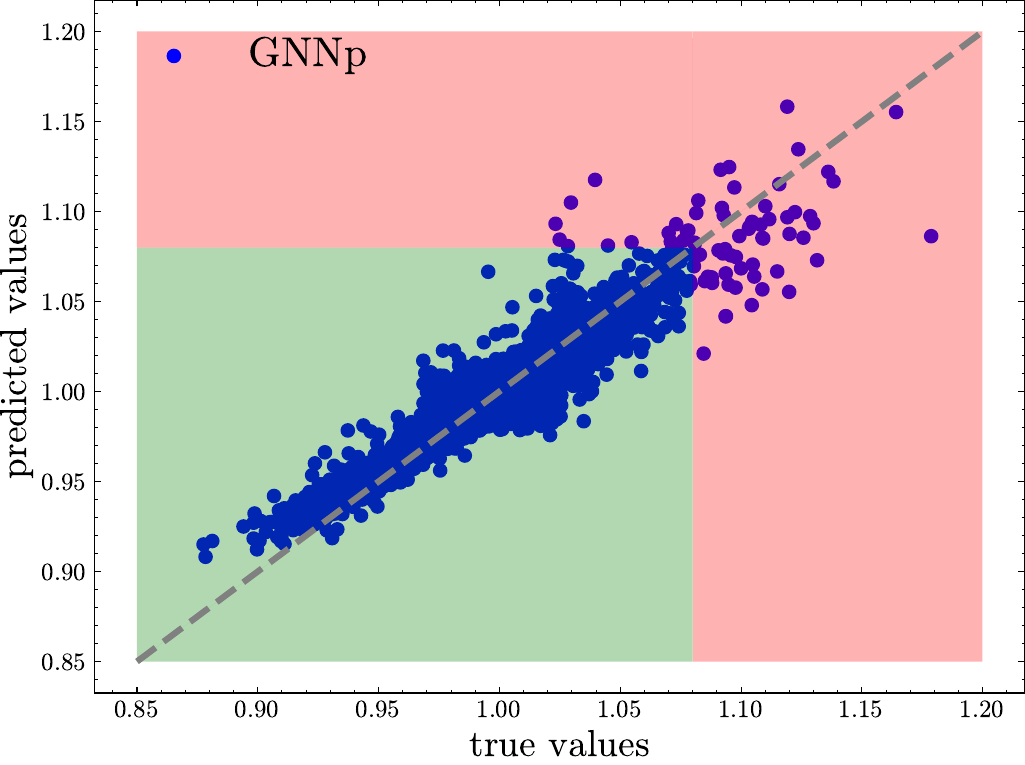}
        \caption{\small{Comparison of GNNp prediction w.r.t. true voltage values.}}
\label{fig:comp1}
    \end{subfigure}
    \hfill
    \begin{subfigure}[b]{0.99\textwidth}
        \centering
        \includegraphics[width=0.8\textwidth]{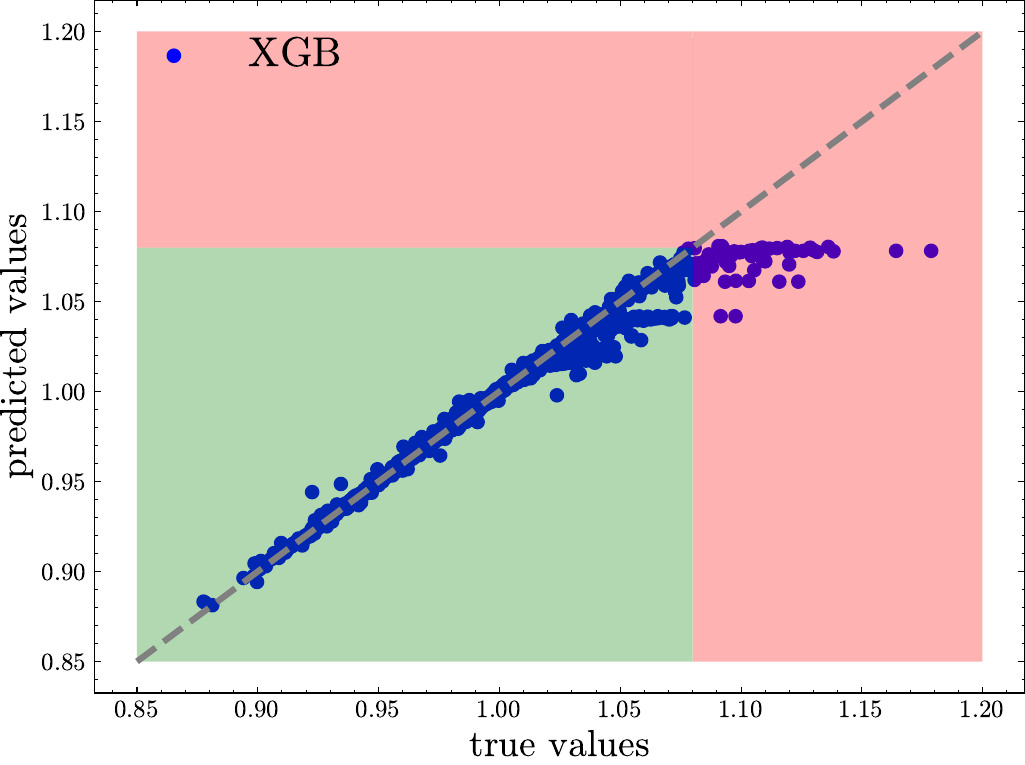}
        \caption{\small{Comparison of XGB prediction w.r.t. true voltage values.}}
\label{fig:comp2}
    \end{subfigure}
    \vspace{-4pt}
    \caption{\small{Comparing predicted and true voltage values. The green zone shows the voltage levels lower than $V_{\max}$ and not seen in the training dataset, while, the red zone shows the out-of-sample scenarios with overvoltage instances.}}
\label{fig:compTot}
\vspace{-4mm}
\end{figure}

The key observations from the numerical results are:
\begin{itemize}
    \item Model-free ACPF fails in predicting instances not seen during the training. On the contrary, the physics-informed ACPF can be used to predict instances not seen by the training data, see Fig. \ref{fig:compTot}. 
    \item The proposed loss function for modelling GNN-based ACPF captures the underlying dynamics better than the benchmark evaluated in this work.
    \item Note for Fig. \ref{fig:voltage} that around 16:30, the GNNp mispredicts a true over-voltage incident. Although the model learned the underlying trend, however, the predicted voltage is below the overvoltage threshold. 
\end{itemize}




 \pagebreak

\section{Conclusion and future work}
\label{section5}


This paper introduces a new physics-informed loss function for training deep neural networks to simulate power flow, incorporating node-wise power imbalance and real power line losses due to heat dissipation. The loss function is used to train a Graph Neural Network (GNNp) on a 3-bus test feeder and is benchmarked against a similar network without physical line losses (GNNb) and a model-free method (XGB).


The proposed model, GNNp, outperforms GNNb on all accuracy metrics due to the inclusion of physical line losses, which prevents training loss divergence. While the model-free XGB shows better overall RMSE and correlation accuracy, this advantage diminishes with larger training sets. XGB fails to predict extreme overvoltage scenarios outside its training set, a limitation of model-free methods. Conversely, GNNp accurately predicts out-of-sample overvoltage trends, indicating it successfully learns the underlying physical model, making it robust and suitable for applications like Hosting Capacity analysis.

In future work, improvements to the current physics-informed model will be tried to be unlocked. 
Additionally, the application of the physical model in multi-phase and large-scale networks will be considered in further research directions. Ultimately, using the physics-informed loss function in an induced-learning fashion to capitalize on combining both in- and out-of-sample prediction capabilities, is a fascinating topic to further explore.



 \pagebreak

\section*{Acknowledgement}
This work is supported by the 
Flemish Government and Flanders Innovation \& Entrepreneurship (VLAIO) through 
IMPROcap project (HBC.2022.0733).
Further, I want to thank my loving parents, brother and sister for supporting me on my academic journeys. Also, a special thanks to my friends at d.a. and BRIKS for keeping me sharp and entertained.

 \pagebreak


\bibliographystyle{IEEEtran}
\bibliography{reference}

\end{document}